\documentclass[preprint,12pt]{elsarticle}
\usepackage{amssymb}
\usepackage{amsmath}
\usepackage{multirow} 
\usepackage{xcolor} 
\usepackage{ulem}
\usepackage[hyphens]{url}
\usepackage[hidelinks]{hyperref}
\hypersetup{breaklinks=true}
\journal{Journal of Quantitative Spectroscopy and Radiative Transfer}
\begin{document}

\begin{frontmatter}

\title{Absorption and scattering properties of nanoparticles in an absorbing medium: modeling with experimental validation}

\author[label1]{Thi Hong Pham}
\author[label1]{Kien Trung Nguyen}
\author[label2]{Viet Tuyen Nguyen}
\author[label1]{Hung Q. Nguyen}
\author[label3]{H. T. M. Nghiem\corref{cor1}}
%% Author affiliation
\affiliation[label1]{organization={Nano and Energy Center, VNU University of Science, Vietnam National University},%Department and Organization
            %addressline={334 Nguyen Trai}, 
            city={Hanoi},
            postcode={120401}, 
            country={Vietnam}}

\affiliation[label2]{organization={Faculty of Physics, VNU University of Science, Vietnam National University},%Department and Organization
            %addressline={334 Nguyen Trai Street, Thanh Xuan Trung Ward, Thanh Xuan District}, 
            city={Hanoi},
            postcode={120401}, 
            country={Vietnam}}

\affiliation[label3]{organization={Phenikaa Institute for Advanced Study, Phenikaa University},%Department and Organization
            %addressline={To Huu Street, Yen Nghia Ward, Ha Dong District}, 
            city={Hanoi},
            postcode={12116}, 
            country={Vietnam}}
%\tnotetext[label4]{hoa.nghiemthiminh@phenikaa-uni.edu.vn}
\cortext[cor1]{hoa.nghiemthiminh@phenikaa-uni.edu.vn}

\begin{abstract}
Absorption and scattering properties of nanoparticles immersed in an absorbing medium are essential in understanding the overall properties of composites and in designing materials with expected functionalities. In this paper, we establish a model based on both Kubelka-Munk theory and Mie theory that links the absorption and scattering properties of individual particles with the reflectance and transmittance spectra of its thin-film composite, supported by detailed experiments. Thin films consisting of TiO$_2$ nanoparticles embedded in PMMA are fabricated on glass substrates using spin-coating and then peeled off to form standalone samples for spectroscopy measurements. By using the Kubelka-Munk theory in combination with the Saunderson correction, the absorption $K$ and scattering $S$ coefficients of multiple nanoparticles are extracted from the measured transmittance and reflectance. 
On the other hand, the absorption $K$ and scattering $S$ coefficients are the sum of absorption and scattering cross-sections of individual particles, which are calculated from the Mie theory specified for particles in an absorbing medium, with the scattering $S$ coefficient further modulated by the anisotropy factor $g$.
The effect of the particulate medium is incorporated through an effective refractive index.
The overall model is validated by matching well between the $K-S$ coefficients extracted from experimental data and theoretical calculations. This agreement provides deep insight into the significant attenuating effect of absorption and scattering on each particle due to the surrounding medium. The validated model of nanoparticles immersed in an absorbing medium can be used to obtain preliminary results for materials designed in a number of applications, such as radiative cooling.
\end{abstract}

\end{frontmatter}

\section{Introduction}
Nanoparticles hold vast applications as being immersed in polymers due to their ability to actively interact with the environments and therefore modify the host properties \cite{schmidt2003properties} such as reinforcing the polymer matrix \cite{odegard2017modeling,guo2013mechanical}, stabilizing the thermal properties together with retarding flame \cite{chrissafis2011can, bikiaris2011can, norouzi2015nanoparticles}, lifting the barrier of polymer to enhance the gas and liquid resistance \cite{cui2015gas, wolf2018shape}, varying the light transmitting in polymer \cite{zhang2017nonlinear,demir2007optical}, preventing the growth of bacteria and other microorganisms \cite{cioffi2005copper,liu2008surface}, changing electrical properties toward electrical conducting, sensing or catalysis \cite{shenhar2005polymer, lu2006synthesis, ma2008effect}. Among those, optical properties can be used for applications in radar \cite{ohlan2008microwave}, sensor \cite{Masheli2023}, metamaterials \cite{pratibha2009tunable, lee2006bioconjugated}, or radiative cooling \cite{caitemperature}. A typical polymer exhibits intense radiation in the mid and far infrared range due to the vibration and stretching of its organic constituents \cite{smith2018infrared}. Recent studies also suggest that polymers absorb in the UV regime \cite{Brissinger19, Hong2023}. In this paper, we are interested in using nanoparticles to modify the optical properties of a polymer in the UV-VIS-NIR regime.

\begin{figure}[htb]
 \centering
    \includegraphics[width=0.8\textwidth]{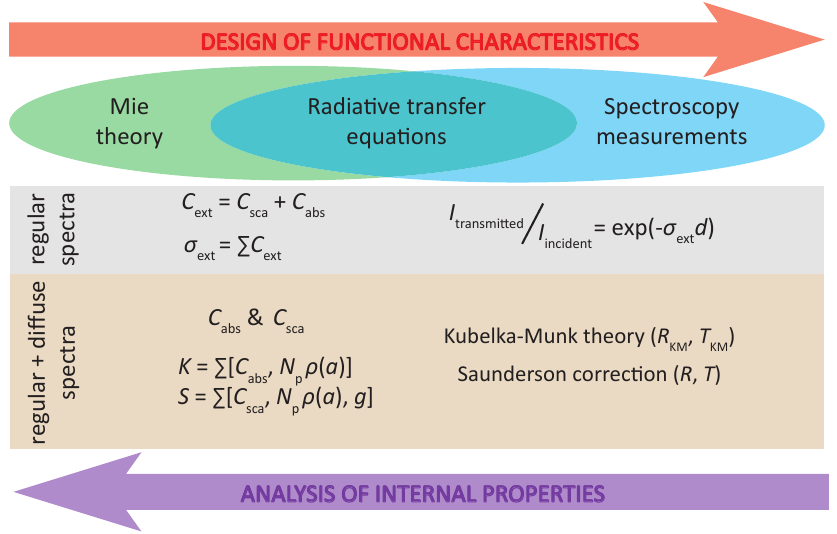}
    \caption{\textbf{The connection between light scattering by an individual particle in a medium and overall optical properties of a thin-film composite} based on Mie theory, radiative transfer equations, and Kubelka-Munk theory. The two-way mapping shows the possibility of component analysis from the spectroscopy measurements, following the purple arrow direction, and the prediction of composite functionality from the calculation of the scattering properties of individual particles in the reversed direction (red arrow).}
\label{fig:schematic}
\end{figure}

\begin{figure}[htb]
    \centering \includegraphics[width=0.8\textwidth]{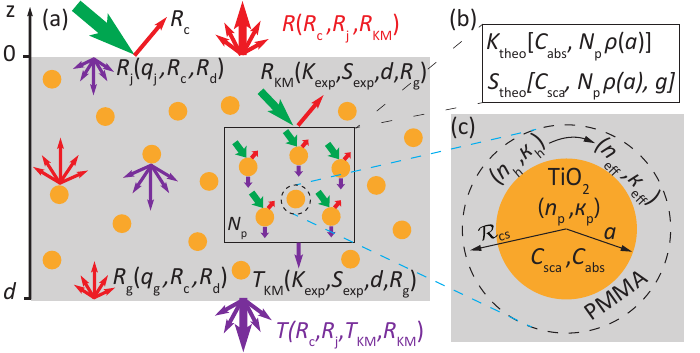}
    \caption{\textbf{Schematic model of for light path in thin-film composite:} (a) Optical paths of light passing through the thin film, where the light is partly absorbed and partly scattered by the composite components. For a given incident light, the total transmittance $T$ (thick purple arrows) passing through the thin-film backside (z = $d$) and total reflectance $R$ (thick red arrows) at the thin-film frontside (z = 0) are measurable in spectroscope experiments with an integrating sphere. The dependence of $T$ and $R$ on the composite components is described in detail in section~\ref{sub:KM} (equations~\eqref{eq:Tkm}-\eqref{eq:R}). Using the equations, the absorption $K_{\rm{exp}}$ and scattering $S_{\rm{exp}}$ coefficients are inverted from the measured $T$ and $R$.
    (b) Given a number of uneven particles per unit volume $N_{\rm{p}}$, the theoretical values $K_{\rm{theo}}$ and $S_{\rm{theo}}$ are defined as the function of absorption $C_{\rm{abs}}$ and scattering $C_{\rm{sca}}$ cross-sections of individual particles, weighted by the particle size distribution $\rho_(a)$ and anisotropy factor $g$ as detailed in section \ref{sub:KM} (equations~\eqref{eq:K_sim}-\eqref{eq:S_sim}).
    (c) Theoretical model for the absorption $C_{\rm{abs}}$ and scattering $C_{\rm{sca}}$ cross-sections of individual particles. Each TiO$_2$ particle with the refractive index $(n_{\rm{p}}, \kappa_{\rm{p}})$ dispersed in PMMA medium with $(n_{\rm{h}}, \kappa_{\rm{h}})$ is assumed to be in a conceptual sphere shown by the dashed line with a radius $\mathcal{R}_{\rm{cs}}$ and an effective refractive-index $(n_{\rm{eff}}, \kappa_{\rm{eff}})$, detail in section~\ref{sub:Mie}}.
    \label{fig:diagram}
\end{figure}

The overall optical properties of thin-film composites with nanoparticles depend on the fundamental optical parameters of their components \cite{volz1975optical}. 
Using the radiative transfer equations, one can mathematically express the variation of light depending on the absorption, scattering, or extinction coefficient. 
Considering the light at the boundaries of the thin film, a procedure is constructed to extract these coefficients from the measured data of transmitted and reflected light, following the step in the backward direction (purple arrow) as shown in Fig.~\ref{fig:schematic}. 
In the case of measuring only regular transmittance, the extinction coefficients of the particles can be obtained, not the absorption and scattering coefficients \cite{book_Quinten2011@chap2, Michael2017}. While, with the case of measuring both the total transmittance and total reflectance, as shown in Fig.~\ref{fig:diagram}a, the $K$ and $S$ coefficients can be extracted within the two-flux Kubleka-Munk theory combined with Saunderson correction \cite{Latour2009, David2013, Wang18, WANG2020}. The total transmittance includes both regular and diffuse transmittance, while the total reflectance comprises both specular and diffuse reflectance.
%, regular and diffuse,

The connection between the Kubelka-Munk coefficients and the optical properties of individual particles is brought up by many studies \cite{Kubelka48, Gate1971, Brinkworth1971, Mudgett_1971, MUDGETT1972, Brinkworth1972, Gate1974, Star1988, Sandoval2014}.
The $K-S$ coefficients depend on both fundamental optical properties of particles in a medium, specifically the absorption coefficient $\mu_{\rm{a}}$ and scattering coefficient $\mu_{\rm{s}}$ per unit path length of the material, and the path length of diffused light. The Kubelka-Munk theory assumes that the scattered light is isotropic, meaning that it is uniformly distributed in all directions. The average path length of the diffused light within a material layer is twice the thickness of that layer. Thus, the absorption coefficient $K$ can be approximated as $K \sim 2\mu_{a}$, this relationship is consistent across several studies \cite{Kubelka48, Gate1971, Brinkworth1971, Mudgett_1971, MUDGETT1972, Brinkworth1972, Gate1974, Star1988, Sandoval2014}.
In early study, the scattering coefficient is treated similarly to the absorption coefficient, i.e., $S \sim 2\mu_{\rm{s}}$ \cite{Kubelka48}. However, the relation between scattering coefficient $S$ and the scattering coefficient per unit path length $\mu_{\rm{s}}$ is not that simple. Comparisons 
between $S$ extracted from the experimental data within the Kubelka-Munk theory and $\mu_{\rm{s}}$ extracted from the same experimental data within the diffusion theory indicate that the $S \approx 3/4 \mu_{\rm{s}}$ \cite{Gate1971}. This approximation is also confirmed through theoretical studies, i.e. the direct connection between diffusion theory and Kubelka-Munk theory \cite{Brinkworth1971}, and the comparison between the results obtained by the two-flux Kubelka-Munk theory and the many-flux theory \cite{Mudgett_1971, MUDGETT1972}.
However, in fact, light diffusion is anisotropic, meaning that scattered light has an uneven angular distribution, leading to a significant influence of the anisotropy factor $g$ on light scattering. Additionally, some studies suggest that the absorption coefficient per unit path also impacts the scattering light. In summary, $S = 3/4(1-g) \mu_{\rm{s}} - x \mu_{\rm{a}}$ with $x=1$ in Ref.~\cite{Brinkworth1972}, $x=0$ in Ref.~\cite{Gate1974}, $x = 1/4$ in Ref.~\cite{Star1988} and $x = 1/4(1-3g)$ in Ref.~\cite{Sandoval2014}. On the other hand, in the limit of small particle density where multiple scattering is negligible, $\mu_{\rm{a}}$ and $\mu_{\rm{s}}$ are simply the sum of the scattering and absorption cross-sections of individual particles. Then for a given number of even particles per unit volume $N_{\rm{p}}$, $\mu_{\rm{a}} = N_{\rm{p}} C_{\rm{abs}}$ and $\mu_{\rm{s}} = N_{\rm{p}} C_{\rm{sca}}$. 
In the case of uneven particles, the expression can be modified accordingly, weighted by the particle size distribution $\rho(\rm{a})$.
%\sout{The additive relations are still valid for the case of dense particles with multiple scattering once the cross-sections are corrected accordingly in the convention of Mie theory, which is discussed below.} 
In general, $K$ and $S$ coefficients can be expressed as a function of the Mie cross-sections, anisotropy factor, particle density and particle size distribution, as shown in Fig.~\ref{fig:diagram}b. From that, the optical properties of the thin-film composite can be predicted from the individual particles, following the step as the right-direction arrow in Fig.~\ref{fig:schematic}.
%. Hence, the anisotropy factor $g$ significantly influences the light scattering, leading to the expression: $S = 3/4(1-g) \mu_{\rm{s}}$ \cite{Brinkworth1972} as demonstrated by comparing with the experimental data \cite{Nobbs1985}. 
%Some studies also suggest that the absorption coefficient per unit path also impacts the scattering light that  $S = 3/4(1-g) \mu_{\rm{s}} - x \mu_{\rm{a}}$ with  $x = 1/4$ in Ref.~\cite{Star1988} and $x = 1/4(1-3g)$ in Ref.~\cite{Sandoval2014}.}

In turn of absorption and scattering at an individual particle, studying the Mie cross-sections is more common in a non-absorbing medium than in an absorbing medium \cite{ Bohren83, Wang2012, Wheeler09}. However, some polymers are shown to absorb strongly in the ultraviolet and infrared regions \cite{smith2018infrared, Brissinger19, Hong2023}, so the contribution of medium absorption to the optical properties of particles must be included.
The absorption and scattering properties of particles in an absorbing medium have been extensively investigated \cite{Mundy1974, Chylekt1977, Bohren1979, Mishchenko2007, Mishchenko2017, MISHCHENKO2018, Quinten1996, Lebedev1999, Sudiarta2001, Fu2001}. 
To solve this problem, in Ref.~\cite{Mundy1974, Chylekt1977}, the authors calculate the extinction, absorption, and scattered energy through a conceptual sphere surrounding a particle. The problem is solved using the far-field approximation in which the conceptual-sphere radius is much greater than the wavelength, and then one can derive the formulas of the corresponding cross-sections.
Other studies also use the far-field approximation to compute the energy received by the detector \cite{Bohren1979, Mishchenko2007, Mishchenko2017, MISHCHENKO2018}, in which the conceptual-sphere radius is the distance from the particle to the detector. 
For the general case, beyond the far-field approximation, as outlined in Ref.~\cite{Quinten1996}, the authors also consider the absorption energy rate through the conceptual-sphere sphere (dashed-line sphere in Fig.~\ref{fig:diagram}c).
%containing a single particle.
%\phamhong{The conceptual sphere with radius $\mathcal{R}_{\rm{cs}}$ represents how light is attenuated by the absorbing medium before being scattered and absorbed by a particle. 
{This conceptual sphere acts as a virtual integrating sphere around each particle to evaluate the energy flow, which does not represent a physical coating. The key physical point is that light is attenuated by the medium before reaching the particle, and the conceptual sphere provides a convenient way to quantify this effect.}
As the result, the traveling waves inside the absorbing medium surrounding the particle, the incident wave and the scattered wave, are shown to be modified with the exponential decay function of the conceptual-sphere radius $\mathcal{R}_{\rm{cs}}$ and the extinction coefficient of the absorbing medium.
%, which has not been fully considered in other studies \cite{Mundy1974, Chylekt1977, Bohren1979, Lebedev1999, Fu2001}. 
With the modified scattered wave and incident wave, the expressions of the extinction and scattered energy rates and the corresponding cross-sections of each particle are straightforward.
Similar expressions of Mie cross-sections are also shown in the other studies \cite{Lebedev1999, Sudiarta2001, Fu2001}; however, the conceptual-sphere radius is set to be equal to the particle radius. The effect of multiple scattering can be incorporated into Mie theory for absorbing media by introducing an effective refractive index that averages the scattering contributions of all particles \cite{Quinten1996}. While this approach neglects near-field interactions between closely spaced particles, as in the generalized Mie theory \cite{Xu1995,Xu1997,Xu1998}, it partially accounts for multiple scattering through increased optical losses at higher particle densities. Therefore, the method is appropriate for estimating bulk optical properties in dilute systems, where interparticle interactions are negligible.

In this work, we establish a model of thin-film composite based on both Kubelka-Munk theory and Mie theory, as shown in the schematic in Fig.~\ref{fig:diagram}. Using this model, we calculate the absorption $K$ and scattering $S$ coefficients of thin-film composites using two approaches: extraction from measured data and calculation from theory as represented by two large arrows in Fig.~\ref{fig:schematic}. In the experimental side, we fabricate thin-film composites consisting of TiO$_2$ nanoparticles embedded in PMMA and then measure their overall reflectance and transmittance, as presented in Fig.~\ref{fig:diagram}a. From these data, the $K$ and $S$ coefficients are extracted using Kubelka-Munk theory and the Saunderson correction following the purple arrow in Fig.~\ref{fig:schematic}. In the theoretical side, the $K$ and $S$ coefficients are the function of the absorption and scattering cross-sections of the individual particles as discussed above and shown in Fig.~\ref{fig:diagram}b. 
In this paper, we calculated the optical properties of single particles using the Mie theory specified for a particle in an absorbing medium with finite values of conceptual-sphere radius as shown in Fig.~\ref{fig:diagram}c. 
In our study of analyzing the absorption and scattering properties of a single particle inside a thin film over a wavelength range of $0.2$ to $2.5$~$\mu$m, the film thickness is on the micrometer scale and so is the conceptual-sphere radius. Therefore, theories based on far-field approximations do not apply in this case, and we use the general theory for an arbitrary conceptual-sphere radius \cite{Quinten1996}.
With that, we also examine the theories developed in Ref.~\cite{Quinten1996} and Refs.~\cite{Lebedev1999, Sudiarta2001, Fu2001} by calculating the absorption and scattering cross-sections of the individual particles in two scenarios: (i) the conceptual-sphere radius in the scale the film thickness \cite{Quinten1996}, and (ii) the conceptual-sphere radius equal to the particle radius \cite{Lebedev1999, Sudiarta2001, Fu2001}.
By comparing the $K-S$ coefficients obtained from the experimental measurements and the theoretical calculations in the two scenarios, we learn that the first scenario yields the better fit than the second scenario. This shows the significant attenuating effect of absorption and scattering on each particle as a result of the surrounding-medium absorption effect. The comparison also validate the overall model. With that, we calculate the optical properties of thin-film composites containing TiO$_2$ particles and air bubbles and discuss these properties in the context of materials applications for radiative cooling.

\section{Experimental and Theoretical Methods} \label{sec:methods}
\subsection{Sample Fabrication and Characterization} \label{sub:fabrication}

TiO$_2$ nanoparticles and PMMA polymer are used to fabricate the thin-film composite. The PMMA obtained from Sumitomo Chemical Co. is in granular form with a molecular weight of 1.18 g/cm$^3$~\cite{Gnanavel19}. TiO$_2$ nanoparticles with purity of 99.8\% are obtained from Shanghai Aladdin Bio-Chem Technology. The sizes of TiO$_2$ particles dispersed in a water medium are measured with a dynamic light scattering apparatus using the ZetaSizer Nano S analyzer (see Appendix~\ref{app1}). The mean and mode values of the particle radius are presented in the second and third columns of table~\ref{tab:distribution}. The mean radii for the two types of TiO$_2$ nanoparticles are 15 nm and 55 nm, denoted as M15 and M55 samples, respectively. Throughout the paper, the corresponding series of samples are labeled as M15 and M55. From the measured data, the particle size distributions are well defined by the log-normal functions with parameters $\mu$ and $\sigma$ determined from the mean and mode values, as shown in table~\ref{tab:distribution}.

\begin{table}
\begin{center}  
  \centering
    \begin{tabular}{|c|c|c|c|c|}
    \hline   
    \rm{Name}&$\rm{mean}$ (nm) & $\rm{mode}$ (nm) & $\mu$ & $\sigma$  \\ \hline
    {M15}  & 15 & 6.75 & 2.44 & 0.72 \\ \hline    
    {M55}  & 55 & 35 & 3.86 & 0.55 \\ \hline      
    \end{tabular}
\end{center}
\caption{\textbf{Parameters of the TiO$_2$ nanoparticle size distribution}: the mean and mode values are measured using the Zetasizer nano S analyzer from Malvern Panalytical. The particle size distribution is described by the log-normal function, $\rho(a) = \frac{1}{a\sigma\sqrt{2\pi}} \exp(-\frac{( \log a-\mu)^2}{2\sigma^2})$, in which $a$ is the particle radius (nm), $\mu$ and $\sigma$ are calculated from the measured mean and mode values.} \label{tab:distribution}
\end{table}

\begin{table}
\begin{center}  
    \begin{tabular}{|c|c|c|c|c|}
    \hline   
    \multirow{2}{*}{Name} & \multicolumn{4}{|c|}{ Average sample thickness $d$ ($\mu$m)}  \\
    \cline{2-5} & $0\%$ $(0\%)$ & $0.5\%$ $(0.14\%)$ &  $1\%$ $(0.28\%)$ & $5\%$ $(1.4\%)$  \\
    \hline
    {series M15}  & 5.1 $\pm$ 0.7  & 5.3 $\pm$ 1.1 & 5.1 $\pm$ 0.8 & 5.5 $\pm$ 0.7 \\ \hline 
    {series M55} &  5.8 $\pm$ 0.9 & 6.7 $\pm$ 1.2 & 8.2 $\pm$ 2.1 & 6.3 $\pm$ 1.1 \\ \hline  
    \end{tabular}
\end{center}
\caption{\textbf{Thickness of eight samples} in series M15 and series M55 with four TiO$_2$ concentrations of mass $c_{\rm{m}}$: 0\%, 0.5\%, 1\%, and 5\%, (the corresponding volume fraction $c_{\rm{v}}$ in parentheses). These values are averaged from five different measurements at various locations in each sample using a KLA Tencor D-100 profilometer.}\label{tab:thickness}
\end{table}

\begin{figure}[htb]
 \centering
    \includegraphics[width=0.8\textwidth]{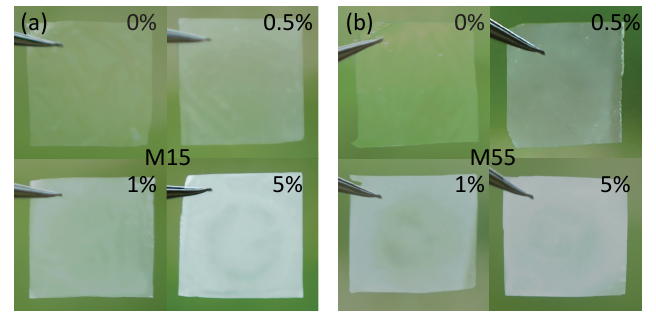}
    \caption{\textbf{Images of thin-film composites of PMMA with different concentrations of TiO$_2$:} 0\%, 0.5\%, 1\%, and 5\% in (a) for series M15 and in (b) for series M55. The sample size is about 2.3 $\times$ 2.3 cm$^2$. The contrast between samples and green (grey) background gradually changes with the increase in particle concentrations.}
      \label{fig:sample}
\end{figure}

Thin-film composites of Poly(Methyl Methacrylate) polymer PMMA with titanium dioxide TiO$_2$ nanoparticles are fabricated in our cleanroom using spin-coating. Initially, PMMA is dissolved in acetone at a ratio of 1:8. After that, TiO$_2$ nanoparticles are added to the mixture, which is then magnetically stirred for 5 hours until uniformly distributed. The mixture of PMMA and TiO$_2$ is spin-coated at 300 rpm in 3 minutes using a standard spin-coater on a glass substrate size 2.54 $\times$ 2.54 cm$^2$. Since acetone evaporates quite fast, the sample dries on itself without any baking steps. Two series of samples M15 and M55 are fabricated at different TiO$_2$ concentrations of mass $c_{\rm{m}}$: 0\%, 0.5\%, 1\%, and 5\%, respectively, in which $c_{\rm{m}}$ presents the ratio of the mass of TiO$_2$ nanoparticles to the mass of the PMMA host. The first set of samples is used to measure the sample thickness using a KLA Tencor D-100 profilometer, as shown in table~\ref{tab:thickness}. These data are used as the input parameters for later calculations, as presented in the following sections. The second set of samples is peeled off from their glass substrates with the help of copper tapes. These standalone films are used for optical measurements, which provide total transmittance and total reflectance data. Being fabricated simultaneously with the identical recipe, these two sets of samples are supposed to have the same morphological structures and optical properties. Images of standalone thin films in series M15 and series M55 are shown in Figs.~\ref{fig:sample}a and~\ref{fig:sample}b, respectively. 

\begin{figure}[htb]
 \centering
    \includegraphics[width=1
    \textwidth]{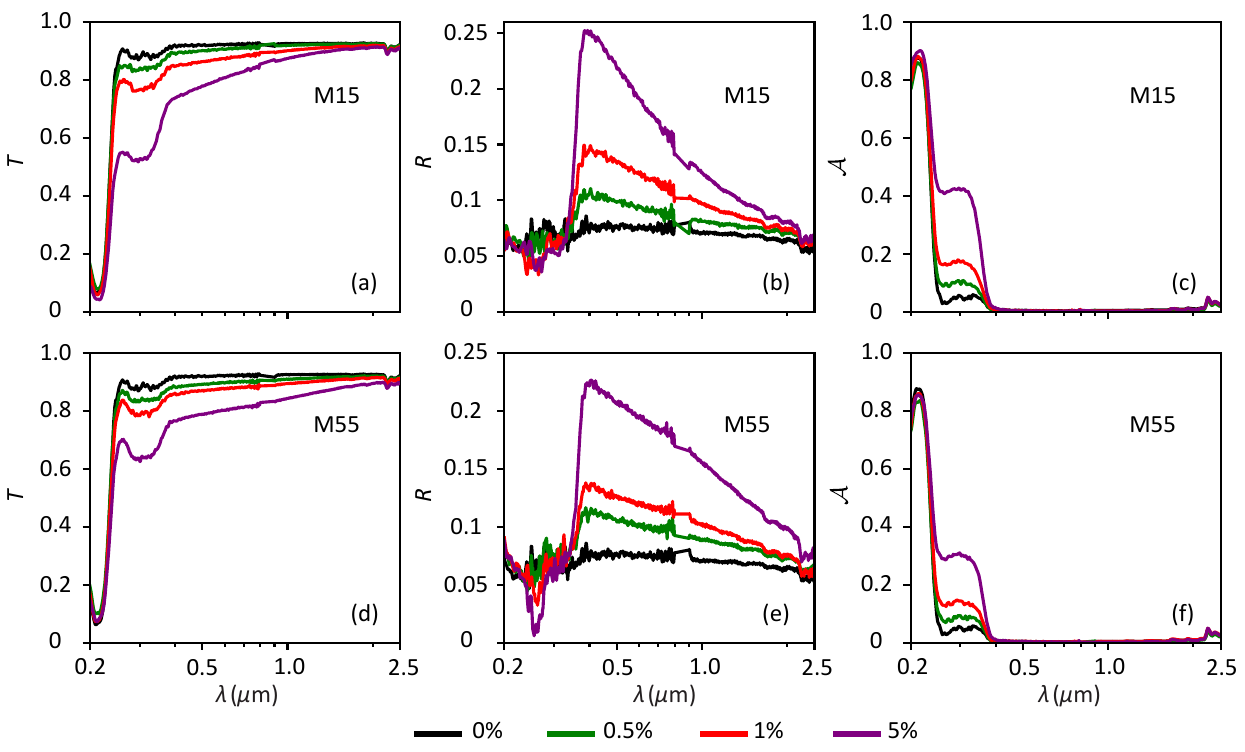}
     \caption{ \textbf{UV-Vis-NIR spectra of thin-film composites.} The total transmittance $T$ is in (a) and (d), and the total reflectance $R$ is in (b) and (e) of series M15 and series M55, respectively. While, the absorption $\mathcal{A}$ is calculated by $\mathcal{A} = 1 - T - R$ in (c) and (f). In each series, there are four samples with different mass concentrations: 0\% in black, 0.5\% in green, 1\% in red, and 5\% in purple. } \label{fig:RTexp}
\end{figure}

The optical properties of thin-film composites are measured in the wavelength from 0.2 to 2.5 $\mu$m. The total transmittance and total reflectance spectra of the standalone thin-film composites are obtained using the Hitachi UH-4150 system with an integrating sphere. We skip data in the region from 0.8 to 0.9 $\mu$m due to a switch in the detector near 0.85 $\mu$m that leads to noisy data. Figs.~\ref{fig:RTexp}a and \ref{fig:RTexp}d present the total transmittance of thin-film composites of series M15 and series M55, respectively, with four different mass concentrations of TiO$_2$ nanoparticles: 0\% in black, 0.5\% in green, 1\% in red, and 5\% in purple. The total reflectance of series M15 and M55 are shown in Figs.~\ref{fig:RTexp}b and~\ref{fig:RTexp}e, respectively. Detailed transmittance and reflectance measurements are presented in our previous report \cite{Hong2023}. From $T$ and $R$, the absorption of the thin-film composites is determined by $\mathcal{A} = 1 - T - R$, as shown in Fig.~\ref{fig:RTexp}c of series M15 and Fig.~\ref{fig:RTexp}f of series M55. The optical properties of both series follow the mass concentrations of TiO$_2$ nanoparticles linearly. 

%%%%
\subsection{The Two-flux Kubelka-Munk Method and Saunderson Correction for Thin-Film Composites} \label{sub:KM}

The two-flux Kubelka-Munk theory is a standard approach that describes light propagation in thin-film composites. The model is formulated using two differential equations for radiation fluxes flowing in both forward and backward directions within the medium \cite{Kubelka48}. The intensities of two radiation fluxes depend on the film thickness $d$, the absorption coefficient $K$, and the scattering coefficient $S$. Solving these equations allows us to determine the transmittance and reflectance of the thin-film composite as shown in Fig.~\ref{fig:diagram}a \cite{Kubelka48, ishimaru1978@book}, which are
\begin{align}
    T_{\rm{KM}} &= \frac {B(1-R_{\rm{g}})}{(A-R_{\rm{g}})\sinh(BSd)+B\cosh(BSd)},\label{eq:Tkm} \\
    R_{\rm{KM}} &= \frac{1-R_{\rm{g}} [A-B\coth(BSd)]}{A-R_{\rm{g}} + B\coth(BSd)},\label{eq:Rkm}
\end{align}
in which $R_{\rm{g}}$ is the backside internal reflectance of the film. $A$ and $B$ are parameters related to the scattering and absorption coefficients
\begin{align}
    A &= \frac{(K+K_{\rm{h}})+S}{S},\\
    B &= \sqrt{A^2 - 1},
\end{align}
in which $K$ and $K_{\rm{h}}$ are the absorption coefficients of particles and host medium respectively. $K_{\rm{h}} = 4\pi \kappa_{\rm{h}}/\lambda$ in which $\kappa_{\rm{h}}$ is the extinction coefficient of the host medium and $\lambda$ is the light wavelength. However, the two-flux Kubelka-Munk theory does not consider the reflection losses at the frontside boundaries. The problem is solved using the Saunderson correction. The total transmittance $T$ and total reflectance $R$, respectively represented by the purple and red arrows at the boundaries in Fig.~\ref{fig:diagram}a, are expressed by \cite{Saunderson1942}
\begin{align}
    T &= \frac {(1-R_{\rm{c}}) T_{\rm{KM}}} {1-R_{\rm{j}} R_{\rm{KM}}},\label{eq:T}\\
    R &= R_{\rm{c}} + \frac{(1-R_{\rm{c}})(1-R_{\rm{j}})R_{\rm{KM}}}{1-R_{\rm{j}} R_{\rm{KM}}}.\label{eq:R}
\end{align}
The reflectances at the frontside boundaries are $R_{\rm{c}}$ and $R_{\rm{j}}$. $R_{\rm{c}}$ indicates the fraction of collimated light reflected at the front interface. $R_{\rm{j}}$ depends on the fraction of light diffused, $q_{\rm{j}}$, inside the materials such that $R_{\rm{j}} = (1 - q_{\rm{j}}) R_{\rm{c}} + q_{\rm{j}} R_{\rm{d}}$, 
in which $R_{\rm{d}}$ indicates the reflectance of diffused light. Similarly, $R_{\rm{g}} = (1- q_{\rm{g}}) R_{\rm{c}} + q_{\rm{g}} R_{\rm{d}}$. The details on the reflectances at the interfaces and the fraction of diffuse light are shown in  Appendices~\ref{app2} and ~\ref{app3}. 
Using equations~(\ref{eq:Tkm}-\ref{eq:R}), we can extract the absorption $K_{\rm{exp}}$ and scattering $S_{\rm{exp}}$ coefficients from the measured transmittance $T$ and reflectance $R$.

As outlined in the introduction, the Kubelka-Munk coefficients are also derived from the absorption and scattering coefficient per unit length and the path length of diffuse light \cite{Kubelka48, Gate1971, Brinkworth1971, Mudgett_1971, MUDGETT1972, Brinkworth1972, Gate1974, Star1988, Sandoval2014}. The absorption coefficient $K$ is defined by the fixed formula, $K = 2\mu_{\rm{a}}$, which equals twice the absorption coefficient per unit length $\mu_{\rm{a}}$ \cite{Kubelka48, Gate1971, Brinkworth1971, Mudgett_1971, MUDGETT1972, Brinkworth1972, Gate1974, Star1988, Sandoval2014}. In contrast, the scattering coefficient $S$ is 
defined by $S = 3/4(1-g) \mu_{\rm{s}} - x \mu_{\rm{a}}$ \cite{Brinkworth1972, Gate1974, Star1988, Sandoval2014}. Since the value of $x$ is not well defined, as mentioned in the introduction, we only take the first contribution to formulate  $S = 3/4(1-g)\mu_{\rm{s}}$.
%The Kubelka-Munk coefficients are also derived from the fundamental optical properties by solving the radiative transfer equation \cite{kubelka-munk1931, Kubelka48}. The absorption coefficient $K$ is defined by the fixed formula, $K = 2\mu_{\rm{a}}$ \cite{Kubelka48, Brinkworth1971, Gate1974, Nobbs1985, Star1988, Sandoval2014}, which equals twice the absorption coefficient per unit length $\mu_{\rm{a}}$. 
%In contrast, the scattering coefficient $S$ is defined by various formulas, as mentioned in the introduction. Here, we use $S = 3/4(1-g)\mu_{\rm{s}}$, which approximates the $S$ coefficient based on the scattering coefficient per unit length $\mu_{\rm{s}}$ and the anisotropy factor $g$ \cite{ Brinkworth1971, Gate1974, Nobbs1985, Star1988, Sandoval2014}.
In the limit of small particle density, light is scattered on a single particle only once, and the chance to be scattered by another is negligible \cite{Bohren2008chap3, book_Quinten2011@chap7, mishchenko@book2002chap3}. Therefore, the $\mu_{\rm{a}}$ and $\mu_{\rm{s}}$ are expressed as the sum of the absorption and scattering cross-section of individual particle \cite{Bohren2008chap3}, which are determined within Mie theory. For a number of even particles $N_{\rm{p}}$, $\mu_{\rm{a}} = N_{\rm{p}} C_{\rm{abs}}$ and $\mu_{\rm{s}} = N_{\rm{p}} C_{\rm{sca}}$. 
To the case of uneven particles with a continuous distribution of particle size $\rho({a})$, the sum of the absorption and scattering cross-sections of particles is modified into an integral form. Therefore, the detailed expression of $K$ and $S$ coefficients shown in Fig.~\ref{fig:diagram}b are formulated by
\begin{align}
     K &= 2 N_{\rm{p}} \int \rho(a) C_{\rm{abs}} {\rm{d}} a, \label{eq:K_sim}\\
     S &= \frac{3}{4} N_{\rm{p}} \int (1-g) \rho({a}) C_{\rm{sca}}  {\rm{d}} a,\label{eq:S_sim}
\end{align}
in which $g$ is the anisotropy factor and $N_{\rm{p}}$ is the particle density, given by
\begin{align}
     N_{\rm{p}} & = \frac{c_{\rm{v}}}{v_{\rm{particle}}} = \frac{c_{\rm{m}} (D_{\rm{h}}/D_{\rm{{p}}})/(1+c_{\rm{m}} D_{\rm{h}}/D_{\rm{p}})}{\int\frac{4}{3}\pi a^3 \rho(a) {\rm{d}}a}, 
     \label{eq:Np}
\end{align}
where $v_{\rm{particle}}$ is the particle volume, $D_{\rm{h}}$ and $D_{\rm{p}}$ are the density of host medium (PMMA) and nanoparticle (TiO$_2$) respectively. 
%The equations are the detailed expression of $K$ and $S$ shown in Fig.~\ref{fig:diagram}b.
%\begin{align}
%     N_{\rm{p}} & = \frac{c_{\rm{v}}}{v_{\rm{particle}}} = \frac{c_{\rm{m}} (D_{\rm{PMMA}}/D_{\rm{{TiO_2}}})/(1+c_{\rm{m}} D_{\rm{PMMA}}/D_{\rm{TiO_2}})}{\int\frac{4}{3}\pi a^3 \rho(a) {\rm{d}}a}, 
%     \label{eq:Np}
%\end{align}
%The absorption and scattering cross-sections of individual particles, $C_{\rm{abs}}$ and $C_{\rm{sca}}$, are calculated using Mie theory. 

Determining the limit of small particle density for single scattering, one can use the condition of volume fraction that is $c_v< 1\%$ or $N_p v_{\text{particle}}< 1$ in the case of even particles \cite{book_Quinten2011@chap7}.  
%However, the scattering cross-section of a particle can also affect the chance of light scattered one more time, therefore the condition is modified into $N_{\rm{p}} C_{\rm{sca}} d \ll 1$ \cite{Bohren2008chap3, mishchenko@book2002chap3} with $d$ the sample thickness. Generalizing for the case of uneven particles and taking into account the anisotropy factor as shown in equation \ref{eq:S_sim}, this condition is then $Sd \ll 1$. We use the scattering coefficient $S$ extracted from experiments to justify if this condition is applied to our case of study, details in section \ref{sub:SKexp}.

%%%%%%%%%%%%%%%%%%%%%%%%555
\subsection{Absorption and Scattering Properties of a Single Particle} \label{sub:Mie}

Inside an absorbing medium, the effect of the surrounding medium on the absorption and scattering coefficients of individual particles is significant, as mentioned in the introduction.
To determine the absorption and scattering properties of a single particle in an absorbing medium, we calculate the cross-sections of the particle with radius $a$.
%\phamhong{The effect of the absorbing medium is represented by a conceptual sphere with arbitrary radius $\mathcal{R}_{\rm{cs}}$ surrounding a particle  \cite{Quinten1996}}.
The effect of the absorbing medium is represented by a conceptual sphere with arbitrary radius $\mathcal{R}_{\rm{cs}}$  \cite{Quinten1996}, illustrated by a dashed line sphere in Fig.~\ref{fig:diagram}c.
The Mie cross-sections for an arbitrary radius of the conceptual sphere are
%The conceptual sphere is illustrated by a dashed line sphere, as shown in Fig.~\ref{fig:diagram}c.
%T\phamhong{ Fig.~\ref{fig:diagram}c describes the theoretical model for calculating the cross-sections through a conceptual sphere with radius $\mathcal{R}_{\rm{cs}}$ surrounding the particle with radius $a$. Here, the conceptual sphere is indicated by a dashed line sphere.}  The Mie cross-sections for an arbitrary effective radius are 
\begin{align}  
    C_{\rm{ext}}  = -&\frac{2\pi}{\lvert k \rvert^2} \sum_{l=1}^{\infty}(2l+1) \label{eq:Cext}\\
    \times&\{ {\rm{Re}}(a_l+b_l) {\rm{Im}}[\xi_l(k\mathcal{R}_{\rm{cs}}) \psi_{l}^{*}{'}(k\mathcal{R}_{\rm{cs}}) - \xi_{l}^{'}(k\mathcal{R}_{\rm{cs}})\psi_{l}^{*}(k\mathcal{R}_{\rm{cs}})]\notag  \\
    & +{\rm{Im}}(a_l + b_l) {\rm{Re}}[\xi_l(k\mathcal{R}_{\rm{cs}}) \psi_{l}^{*}{'}(k\mathcal{R}_{\rm{cs}}) - \xi_{l}^{'}(k\mathcal{R}_{\rm{cs}})\psi_{l}^{*}(k\mathcal{R}_{\rm{cs}})] \notag  \\
    & +\frac{{\rm{Im}}(k)}{{\rm{Re}}(k)} {\rm{Re}}(a_l - b_l) {\rm{Re}}[\xi_l (k\mathcal{R}_{\rm{cs}}) \psi_{l}^{*}{'}(k\mathcal{R}_{\rm{cs}}) + \xi_{l}^{'}(k\mathcal{R}_{\rm{cs}}) \psi_{l}^{*}(k\mathcal{R}_{\rm{cs}})]\notag  \\
    & -\frac{{\rm{Im}}(k)}{{\rm{Re}}(k)} {\rm{Im}}(a_l - b_l) {\rm{Im}}[\xi_l (k\mathcal{R}_{\rm{cs}}) \psi_{l}^{*}{'}(k\mathcal{R}_{\rm{cs}}) +\xi_{l}^{'}(k\mathcal{R}_{\rm{cs}}) \psi_{l}^{*}(k\mathcal{R}_{\rm{cs}})]\},\notag  \\
    C_{\rm{sca}}  = - & \frac{2\pi}{\lvert k \rvert^2} \sum_{l=1}^{\infty}(2l+1)\times \{(\lvert a_l\rvert ^2 + \lvert b_l\rvert ^2) {\rm{Im}}[\xi_l(k\mathcal{R}_{\rm{cs}}) \xi_{l}^{*}{'}(k\mathcal{R}_{\rm{cs}})] \notag \\
    & + \frac{{\rm{Im}}(k)}{{\rm{Re}}(k)}(\lvert a_l\rvert ^2 - \lvert b_l\rvert ^2) {\rm{Re}}[\xi_l(k\mathcal{R}_{\rm{cs}}) \xi_{l}^{*}{'}(k\mathcal{R}_{\rm{cs}})]\}, \label{eq:Csca}  \\
    C_{\rm{abs}}  = \quad & C_{\rm{ext}} - C_{\rm{sca}}, \label{eq:Cabs}
\end{align}
where $\psi_l$ and $\xi_l$ are the Riccati-Bessel functions. $k=2\pi(n_{\rm{h}} + i\kappa_{\rm{h}})/\lambda$, in which $(n_{\rm{h}} + i\kappa_{\rm{h}})$ is the refractive index of the host medium. The two parameters $a_l$ and $b_l$ are Mie coefficients \cite{Bohren83}, 
%$l$ is the multipole expansion.
\begin{align}
    a_l & =\frac{m\psi_l(mx)\psi_l'(x)-\psi_l(x)\psi_l'(mx)}{m\psi_l(mx)\xi_l'(x)-\xi_l(x)\psi_l'(mx)}, \label{eq:an} \\ 
    b_l & =\frac{\psi_l(mx)\psi_l'(x) - m\psi_l(x)\psi_l'(mx)}{\psi_l(mx)\xi_l'(x) - m\xi_l(x)\psi_l'(mx)}, \label{eq:bn}
\end{align}
%\hn{COMMENTS: check the equations here. Are they correct?}
where $x = ka$ is the size parameter, $m = (n_{\rm{p}} + i\kappa_{\rm{p}})/(n_{\rm{h}} + i\kappa_{\rm{h}})$, and $(n_{\rm{p}} + i\kappa_{\rm{p}})$ is the refractive index of a particle. 
Thin-film composite, including the particles immersed in a polymer medium, becomes a particulate environment with the refractive index $(n_{\rm{eff}}+i\kappa_{\rm{eff}})$ \cite{William1997}. It represents the mean effect of other particles inside the composite on a given particle (see Appendix~\ref{app4}).  
Thus, we replace $(n_{\rm{h}} + i \kappa_{\rm{h}})$ by $(n_{\rm{eff}} + i \kappa_{\rm{eff}})$ in equations~\eqref{eq:Cext}-\eqref{eq:bn} as using the Mie cross-sections of a single particle to calculate the total absorption and scattering coefficients of all the particles, equations~\eqref{eq:K_sim} and~\eqref{eq:S_sim}. Numerical calculations were carried out using our own Python code. The implementation of Mie scattering calculations follows the classic BHMIE algorithm \cite{Bohren83}, with modifications to the absorption and scattering cross sections described above.

Other studies~\cite {Lebedev1999,Sudiarta2001,Fu2001} also examine the optical properties of single particles in an absorbing medium with similar expressions as in Ref.~\cite{Quinten1996}; however, the conceptual radius is equal to the particle radius. {To evaluate these theories, we calculate the Mie cross-sections in the two limits of $\mathcal{R}_{\rm{cs}}$: $\mathcal{R}_{\rm{cs}} = d/2$ at maximum and $\mathcal{R}_{\rm{cs}} = a$} at minimum. From those, the corresponding Kubelka-Munk coefficients $S$ and $K$ are compared with those extracted from experimental data. 

%\phamhong{Our theoretical calculations are performed using our own Python-based code, based on the classic BHMIE algorithm \cite{Bohren83}. The expressions for the absorption and scattering cross-sections are modified according to equations~\eqref{eq:Cext}-\eqref{eq:bn} to be suitable for particles embedded in an absorbing medium.}

\section{Results and Discussion} \label{Sec:results}
%\subsection{Backward analysis of the inside properties} \label{sub:SKresults}

\subsection{$K$-$S$ Extracted from Experiments} \label{sub:SKexp}

\begin{table}
\begin{center}  
    \begin{tabular}{|c|c|c|c|}
    \hline   
    \multirow{2}{*}{Name} & \multicolumn{3}{|c|}{ Diffuse light fraction $q_{\rm{g}}$}  \\
    \cline{2-4} & $0.5\%$ &  $1\%$ & $5\%$  \\ \hline
    {series M15} &  0.127 & 0.245 & 0.5 \\ \hline  
    {series M55}  &  0.069 & 0.169 & 0.5 \\ \hline      
    \end{tabular}
\end{center}
\caption{ \textbf{Fraction of light diffused at front interface $q_{\rm{g}}$} with three concentrations: 0.5\%, 1\% and 5\% of series M15 and M55. We assume that the fraction of light diffused at the back interface $q_{\rm{j}}$ = $2 q_{\rm{g}}$, which imposes the value of $q_{\rm{g}} \leq 0.5$.}\label{tab:q}
\end{table}

%\hn
{To determine the $K_{\rm{exp}}$ and $S_{\rm{exp}}$ coefficients of particles from measured data with equations~\eqref{eq:Tkm}-\eqref{eq:R}, we need to know the internal reflectance of the film at both interfaces $R_{\rm{j,g}}$, while the $R_{\rm{j,g}}$ determination requires the fractions of diffused light $q_{\rm{j,g}}$  (see in Appendix~\ref{app2}). Following the investigation of the possible values of $q_{\rm{j,g}}$ in Appendix~\ref{app3}, we assume $q_{\rm{g}}$ as the total particle cross-section in the unit of sample area and $q_{\rm{j}}=2q_{\rm{g}}$. The values of $q_{\rm{g}}$ for each series with the different concentrations are shown in table~\ref{tab:q}. These values fall within the reasonable ranges of $q_{\rm{g}}$ and $q_{\rm{j}}$, the light areas in Fig.~\ref{fig:heatmap}.}
\begin{figure}[htb]
 \centering
    \includegraphics[width=1\textwidth]{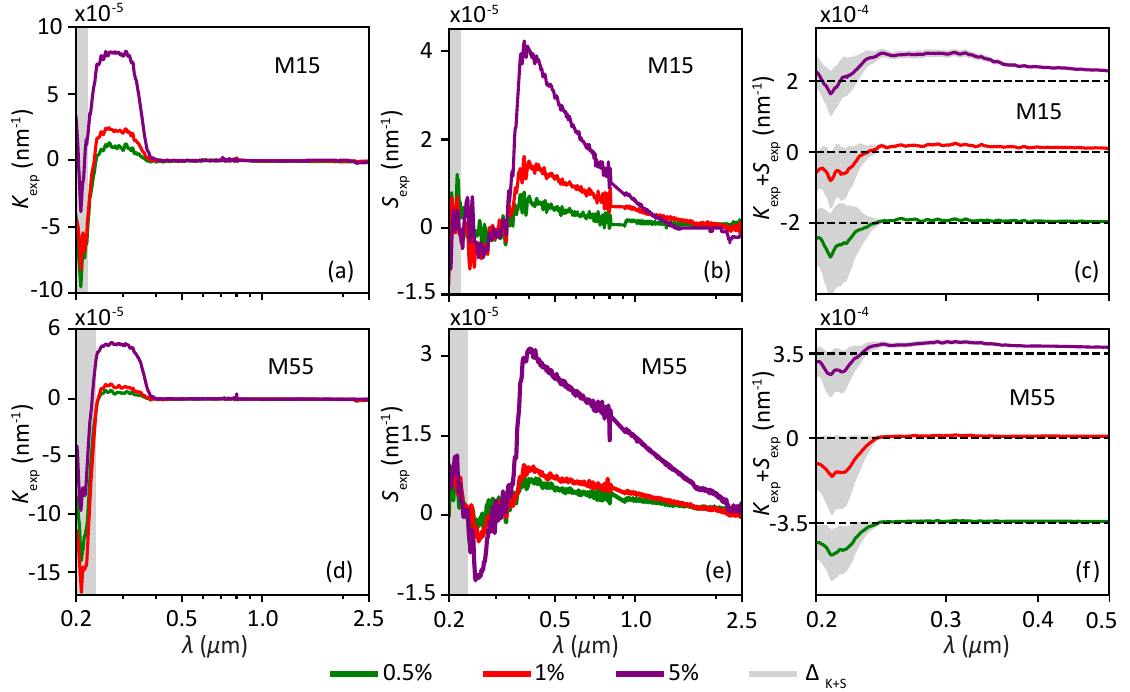}
     \caption{\textbf{(Color online) $K_{\rm{exp}}$-$S_{\rm{exp}}$ coefficients extracted from measured data:} series M15 in (a) and (b), series M55 in (d) and (e). The $K_{\rm{exp}}$-$S_{\rm{exp}}$ values are determined by inverting equations~\eqref{eq:Tkm}-\eqref{eq:R} in the wavelength range from 0.2 to 2.5 $\mu$m with three mass concentrations of 0.5\% in green, 1\% in red, and 5\% in purple, respectively. The uncertainty of $(K_{\rm{exp}}+S_{\rm{exp}})$ in gray area determined by equation \eqref{eq:errorSK} is presented in (c) for series M15 and in (f) for series M55.}
      \label{fig:SKexp}
\end{figure}

The absorption coefficient $K_{\rm{exp}}$ of TiO$_2$ nanoparticles extracted from measured data in series M15 and series M55 are shown in Figs.~\ref{fig:SKexp}a and \ref{fig:SKexp}d with three mass concentrations: 0.5\% in green, 1\% in red, and 5\% in purple, respectively. Since TiO$_2$ strongly absorbs in the ultraviolet region (see Fig.~\ref{fig:nk_eff}b in Appendix~\ref{app4}), the absorption coefficient $K_{\rm{exp}}$ of TiO$_2$ nanoparticles linearly increases with the increasing mass concentrations of TiO$_2$ in the wavelength range of $0.24-0.38$~$\mu$m, while they are approximately zero in the visible and near-infrared region. However, $K_{\rm{exp}}$ strongly decays at $\lambda < 0.24$~$\mu$m due to the strong absorption of the medium in this range. This point is made clear by comparing it to the theoretical results in the following section. Figs.~\ref{fig:SKexp}b and \ref{fig:SKexp}e respectively present the scattering coefficient $S_{\rm{exp}}$ of series M15 and series M55. The scattering coefficient $S_{\rm{exp}}$ strongly depends on the increased mass concentration in the visible and near-infrared regions and weakly changes in the ultraviolet region.

In extracting $K_{\rm{exp}}$ and $S_{\rm{exp}}$ coefficients of TiO$_2$ nanoparticles, the accuracy of our final results depends on the sample and the optical measurements. In this paper, it is mainly determined by the sample thickness since the uncertainty of the thickness is over $10\%$ while that of the optical measurements is below $1\%$. Analysing the uncertainty of the total $(K_{\rm{exp}} + S_{\rm{exp}})$ from equations~\eqref{eq:Tkm} and~\eqref{eq:Rkm} \cite{Hughes10}, we have
\begin{align}
      \Delta_{K+S} \approx \frac{K_{\rm{h}} \Delta d}{d}, \label{eq:errorSK}
\end{align}
where $\Delta d$ is the error in thickness measurement as shown in table~\ref{tab:thickness}, $K_{\rm{h}}$ is the absorption coefficient of the PMMA host medium.
In Figs.~\ref{fig:SKexp}c and \ref{fig:SKexp}f, the total value  $(K_{\rm{exp}} + S_{\rm{exp}})$ is showed with uncertainty $\pm \Delta_{K+S}$ highlighted in gray. 
Similar to the discussion in Appendix~\ref{app3}, to keep the physical meaning of dissipating and redirecting of light within the Kubelka-Munk theory, $K$-$S$ coefficients are restricted to be non-negative \cite{Kubelka48}. Therefore, from the points where $(K_{\rm{exp}} + S_{\rm{exp}} \pm \Delta_{K+S})$ is negative, we define the unreliable ranges, in particular $\lambda \approx 0.2 - 0.22$ $\mu$m for series M15, and $\lambda \approx 0.2 - 0.234$ $\mu$m for series M55. We use gray rectangles to annotate the ranges in Figs.~\ref{fig:SKexp}a,~\ref{fig:SKexp}b,~\ref{fig:SKexp}d, and~\ref{fig:SKexp}e and the following figures. These ranges almost match with the range of unphysically negative $K_{\rm{exp}}$ and $S_{\rm{exp}}$ in the figures. One may improve the extraction of $K_{\rm{exp}}$ and $S_{\rm{exp}}$ by enhancing the flatness of the sample.

\subsection{$K$-$S$ Calculated from Theory}
\label{sub:SKtheory}

The theoretical values of $K_{\rm{theo}}$ and $S_{\rm{theo}}$ are calculated following equations~(\ref{eq:K_sim}-\ref{eq:S_sim}). As  mentioned in section \ref{sub:KM}, equations (\ref{eq:K_sim}) and (\ref{eq:S_sim}) are used with the condition of single scattering $c_v<1$. This limit, however, can be relaxed in lossy media, where increased particle density enhances the effective absorption, thereby reducing the strength of multiple scattering interaction, as demonstrated by other studies~\cite{Durant2007, Durant2007@Monte, Zhai2023@clusters}. Therefore, the equations can be applied to our study case, where the effective refractive index is used in the calculation of Mie cross-sections of the individual particles.
With the effective refractive index of the host medium $(n_{\rm{eff}}+i \kappa_{\rm{eff}})$, and the refractive index of particle $(n_{\rm{p}}+i \kappa_{\rm{p}})$, the absorption $C_{\rm{abs}}$ and scattering $C_{\rm{sca}}$ cross-sections are calculated using equations~\eqref{eq:Cext}-\eqref{eq:bn}. The values of these refractive indices are shown in Fig.~\ref{fig:nk_eff} in Appendix~\ref{app4}. 
In order to justify the effect of the surrounding medium on the absorption and scattering cross-sections of a single particle, we calculate the Mie cross-sections in the two scenarios: $\mathcal{R}_{\rm{cs}} = d/2$ \cite{Quinten1996} and $\mathcal{R}_{\rm{cs}} = a$ \cite{Lebedev1999, Sudiarta2001, Fu2001}.
The theoretical values of $K_{\rm{theo}}$ and $S_{\rm{theo}}$ are then compared with $K_{\rm{exp}}$ and $S_{\rm{exp}}$ obtained in the previous section.
%we calculate the $K_{\rm{theo}}$ and $S_{\rm{theo}}$ coefficients of TiO$_2$ nanoparticles in PMMA medium from the Mie cross-sections, $C_{\rm{abs}}$ and $C_{\rm{sca}}$. 
\begin{figure}[htb]
 \centering
 \includegraphics[width=0.8\textwidth]{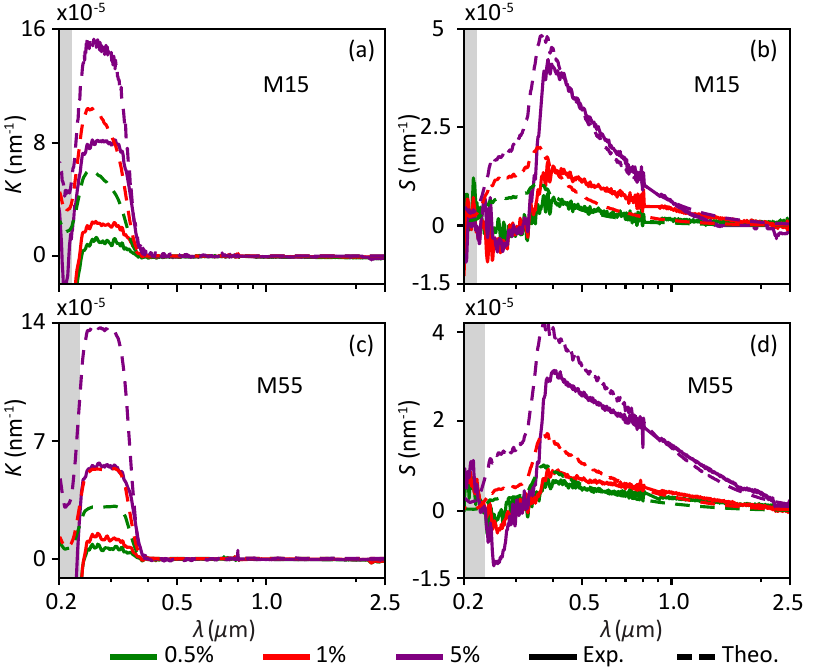}
     \caption{\textbf{(Color online) Comparison between theoretical and experimental results $(K,S)$ in the $\mathcal{R}_{\rm{cs}} = d/2$ limit} ($d$ is the sample thickness): series M15 in (a) and (b), series M55 in (c) and (d). The theoretical values of $K_{\rm{theo}}$ and $S_{\rm{theo}}$ depicted by the dashed line are computed using equations~\eqref{eq:K_sim}-\eqref{eq:bn} with $\mathcal{R}_{\rm{cs}} = d/2$. The experimental data $(K_{\rm{exp}}$, $S_{\rm{exp}})$ shown by the solid line are the same as those in Fig.~\ref{fig:SKexp}. Data corresponding to three mass concentrations are shown in green for 0.5\%, red for 1\%, and purple for 5\%.
     }  \label{fig:SKsim_R=d/2}
\end{figure}

In the first scenario $\mathcal{R}_{\rm{cs}} = d/2$, the calculated $K_{\rm{theo}}$ and $S_{\rm{theo}}$ of TiO$_2$ nanoparticles in PMMA are presented as dashed lines in Fig.~\ref{fig:SKsim_R=d/2}, while and $K_{\rm{exp}}$ and $S_{\rm{exp}}$ coefficients of TiO$_2$ nanoparticles are shown as solid lines. Figures on the top row are $K$ and $S$ coefficients of series M15 with three mass concentrations, while figures on the bottom row are those from series M55. In this scenario, 
{the calculated $K_{\rm{theo}}$ coefficients in Figs.~\ref{fig:SKsim_R=d/2}a and ~\ref{fig:SKsim_R=d/2}c show a similar trend to the experimental data of the corresponding series M15 and series M55.}
The values of the absorption coefficient depend on the concentration and type of TiO$_2$ nanoparticles together with the characterization of PMMA at different wavelengths. PMMA exhibits the highest absorption at~$\lambda = 0.22$~$\mu m$, as presented in Fig.~\ref{fig:nk_eff}a, Appendix \ref{app4}. As a result, the absorption coefficients $K_{\rm{theo}}$ of the total TiO$_2$ particles strongly decay, which is in contrast to the strong absorption properties of TiO$_2$ in bulk form in the corresponding range, as shown in Fig.~\ref{app4}b obtained to Ref.\cite{Siefke2016}. This means that the absorbing medium strongly affects the $K$ coefficient of the TiO$_2$ nanoparticles. This effect is also demonstrated in the $S_{\rm{theo}}$ coefficient spectra in the ultraviolet region in Fig.~\ref{fig:SKsim_R=d/2}b of series M15 and Fig.~\ref{fig:SKsim_R=d/2}d of series M55. 
Similar to the behavior of $K_{\rm{theo}}$ coefficient, the tendency of the $S_{\rm{theo}}$ coefficient is also compatible with those of the experimental data. The magnitudes of the scattering coefficient calculated from the theory in series M15 are nearly equal to those extracted from experiments. However, the results of $S_{\rm{theo}}$ and $S_{\rm{exp}}$ in series M55 do not match each other as well as those in series M15.
It is due to the precision of the particle size measurement limited to large particles \cite{BHATTACHARJEE@DLS}, then the input data of the particle size distribution in series M55 in the theoretical calculation are less accurate than those in series M15. Despite these differences, the calculations with $\mathcal{R}_{\rm{cs}} = d/2$ produce $K_{\rm{theo}}$ and $S_{\rm{theo}}$ coefficients with values that closely match the scale and trend of the measurement data. 
%In the scattering spectra, the magnitudes calculated from theory and extracted from experiments in series M15 match better than those from series M55 due to the accuracy of the measurement of particle sizes. This is because the large particles settle to the bottom cuvette more quickly than the smaller ones, making it difficult to collect the correct data.
%Similarly, the $S_{\rm{theo}}$ coefficient calculated from the theory and extracted from the experiment match well, as shown in Fig.~\ref{fig:SKsim_R=d/2}b of series M15 and Fig.~\ref{fig:SKsim_R=d/2}d of series M55. The magnitude of $K$-$S$ between the theory and experiment of series M55 differs more than that of series M15. This is because the large particles settle to bottom more quickly than small ones, so they can not be inspected by the laser light source in the particle size measurement. Although the accuracy measurement of input data affects the theoretical results, calculations based on Ref.~\cite{Quinten1996} produce $K_{\rm{theo}}$ and $S_{\rm{theo}}$ with values closely matching the measurement data. 
%The optical properties of the TiO$_2$ nanoparticles depend on the concentration and type of TiO$_2$ particles together with the characterization of PMMA at different wavelengths. Since PMMA has the highest absorption at $\lambda = 0.22$ $\mu m$  (presented in Fig.~\ref{fig:nk_eff}a, Appendix \ref{app4}), it makes the absorption coefficient $K_{\rm{theo}}$ of TiO$_2$ nanoparticles no full absorption as TiO$_2$ bulk at $\lambda \leq 0.24$ $\mu$m.
\begin{figure}[htb]
 \centering
 \includegraphics[width=0.8\textwidth]{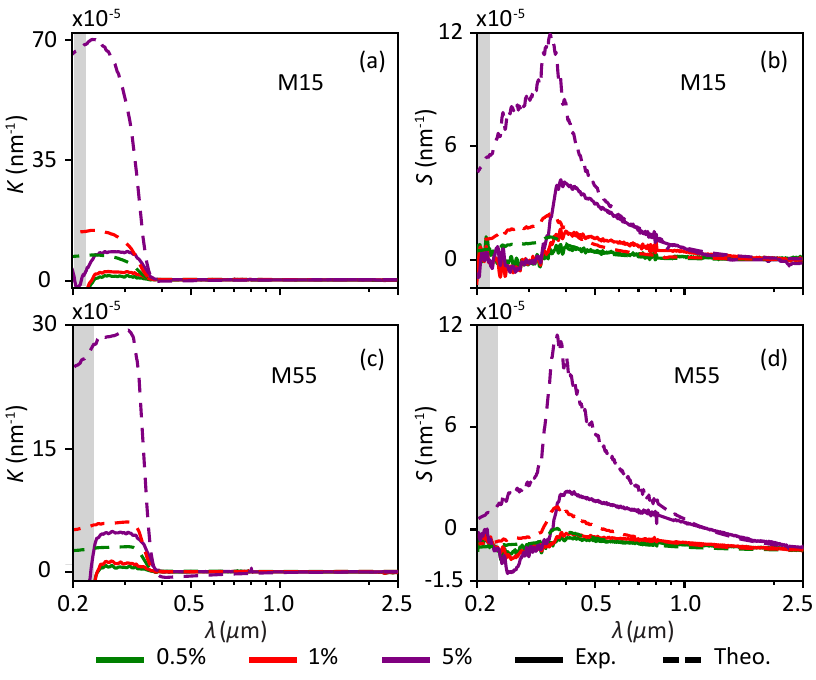}
     \caption{\textbf{(Color online) Comparison between theoretical and experimental results $(K,S)$ in the $\mathcal{R}_{\rm{cs}} = a$ limit} ($a$ is the particle radius): series M15 in (a) and (b), series M55 in (c) and (d). The theoretical values of $K_{\rm{theo}}$ and $S_{\rm{theo}}$ depicted by the dashed line are computed using equations~\eqref{eq:K_sim}-\eqref{eq:bn} with $\mathcal{R}_{\rm{cs}} = a$. The experimental data $(K_{\rm{exp}}$, $S_{\rm{exp}})$ shown by the solid line are the same as those in Fig.~\ref{fig:SKexp}. Data corresponding to three mass concentrations are shown in green for 0.5\%, red for 1\%, and  purple for 5\%.} \label{fig:SKsim_dr=a}
     %The $K$-$S$ of immersed particles is also calculated by equations~\ref{eq:K_sim}-~\ref{eq:bn}. The optical properties of immersed particles are calculated equivalent to the theory in Ref.~\cite{Lebedev1999}.} \label{fig:SKsim_dr=a}
\end{figure}

In the second scenario, the $K_{\rm{theo}}$ and $S_{\rm{theo}}$ coefficients are calculated in the limit of the conceptual radius equals to the particle radius $\mathcal{R}_{\rm{cs}} = a$. Similar to the case $\mathcal{R}_{\rm{cs}} = d/2$, we compute the $K_{\rm{theo}}$ and $S_{\rm{theo}}$ coefficients for series M15 and series M55 as shown in dashed lines in Fig.~\ref{fig:SKsim_dr=a}. These theoretical results are compared to experimental data as shown in solid lines. The $K_{\rm{theo}}$ coefficients of series M15 in Fig.~\ref{fig:SKsim_dr=a}a and series M55 in Fig.~\ref{fig:SKsim_dr=a}c are approximately zero coinciding with the experimental data in the visible and near-infrared regions due to the non-absorbing properties of TiO$_2$ nanoparticles and the PMMA medium. However, the theoretical $K_{\rm{theo}}$ spectra do not exhibit the tendency and the scale of the experimental data in the ultraviolet regime. Likewise, the theoretical $S_{\rm{theo}}$ spectra of series M15 in Fig.~\ref{fig:SKsim_dr=a}b and series M55 in Fig.~\ref{fig:SKsim_dr=a}d do not match with experimental results in all regions. Therefore, the theory of light absorption and scattering on particles with $\mathcal{R}_{\rm{cs}} = a$~\cite{Lebedev1999} does not truly describe the effect of the absorbing medium. In conclusion, with these comparisons of the theoretical results and the experimental data, the first scenario $\mathcal{R}_{\rm{cs}} = d/2$ appears to be better than the second one $\mathcal{R}_{\rm{cs}} = a$. Our calculations also suggest that absorption and scattering on each particle are strongly attenuated by the strong absorption of the surrounding medium. 
%This matchs well with the recent observation of reduction on the localized surface plasmon resonance due to the absorbing medium \cite{Masheli2023}.the weak effect of the absorbing medium presented by

%%%%%%
\subsection{Prediction of the Overall Properties of Thin-Film Composite} \label{sub:forward}
\begin{figure}[htb]
 \centering
 \includegraphics[width=1\textwidth]{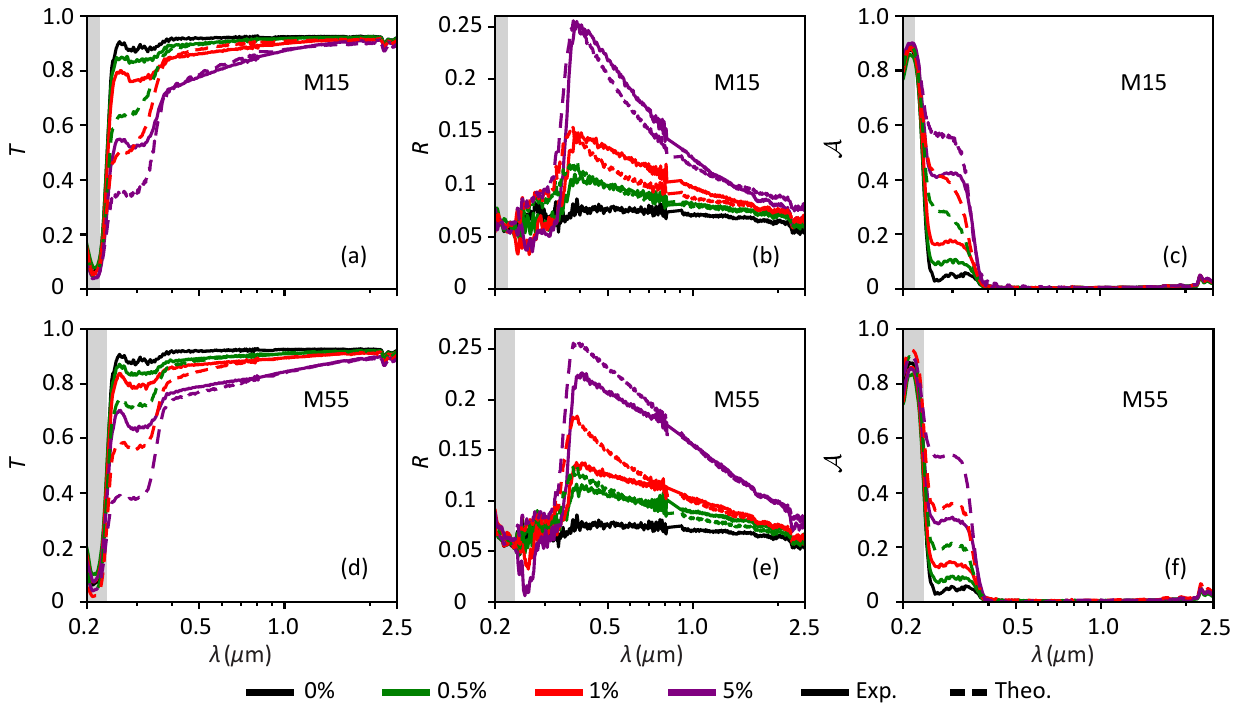}
     \caption{\textbf{(Color online) Overall optical properties of thin-film composites (TiO$_2$/PMMA):} series M15 in (a)-(c), while series M55 in (d)-(f). From the absorption $K$ and scattering $S$ coefficients calculated from theory with $\mathcal{R}_{\rm{cs}}=d/2$ in Fig.~\ref{fig:SKsim_R=d/2}, the total transmittance $T$ and total reflectance $R$ are calculated straightforwardly following equations~\eqref{eq:Tkm}-\eqref{eq:R}, while the absorption $\mathcal{A}$ is determined by $\mathcal{A} = 1 - T -R$. The calculated results are shown in the dashed lines, while the experimental data in the solid lines are the same as those in Fig.~\ref{fig:RTexp}. }    \label{fig:RTtheo_d2}
\end{figure}
In section~\ref{sub:SKtheory}, we demonstrate the effect of the absorbing medium on the optical properties of immersed nanoparticles with the reasonable use of $\mathcal{R}_{\rm{cs}} = d/2$. With these theoretical $K_{\rm{theo}}$-$S_{\rm{theo}}$ results fitted well to the experimental data, we calculate the overall optical properties of the thin-film composite following equations~\eqref{eq:Tkm}-~\eqref{eq:R}. 
Despite potential numerical precision issues in computing $C_{\rm{abs}}$ via equation~\eqref{eq:Cabs} when $C_{\rm{ext}} \approx C_{\rm{sca}}$ in the wavelength range 0.38 - 2.5 $\mu$m, the distinct physical regimes are preserved. Notably, significant $K$ values in the ultraviolet (0.2–0.38 $\mu$m) indicate robust absorption, while diminished $K$ values at longer wavelengths (0.38–2.5 $\mu$m) correspond to reduced absorption. Importantly, these minor precision issues do not impact the macroscopic properties ($R$, $T$, $\mathcal{A}$).
Calculations based on our theoretical model produce $T$, $R$, and $\mathcal{A}$ with values that match well with measured data presented in Fig.~\ref{fig:RTtheo_d2}. 

For the radiative cooling application in the wavelength regime $\lambda~\sim 0.3-2.5$ $\mu$m (solar radiation spectra), a material is expected to meet two requirements: high reflectance, and low absorption \cite{passivecooling@glass-polymer2017, passivecooling@science2018}.
%mid-infrared emittance in the atmosphere transmission window ($\lambda \sim 8-13$ $\mu$m) 
Even though TiO$_2$ is commonly used as pigments in reflective paints, it satisfies only a part of the first requirement as showing the enhancement of reflection in the range from $\lambda \sim 0.38-2.5$~$\mu$m, as in Figs.~\ref{fig:RTtheo_d2}b and~\ref{fig:RTtheo_d2}e.
However, its strong absorption in the UV range, Figs.~\ref{fig:RTtheo_d2}c and~\ref{fig:RTtheo_d2}f, may negatively affect the performance of the material.

\begin{figure}[htb]
 \centering
 \includegraphics[width=1 \textwidth]{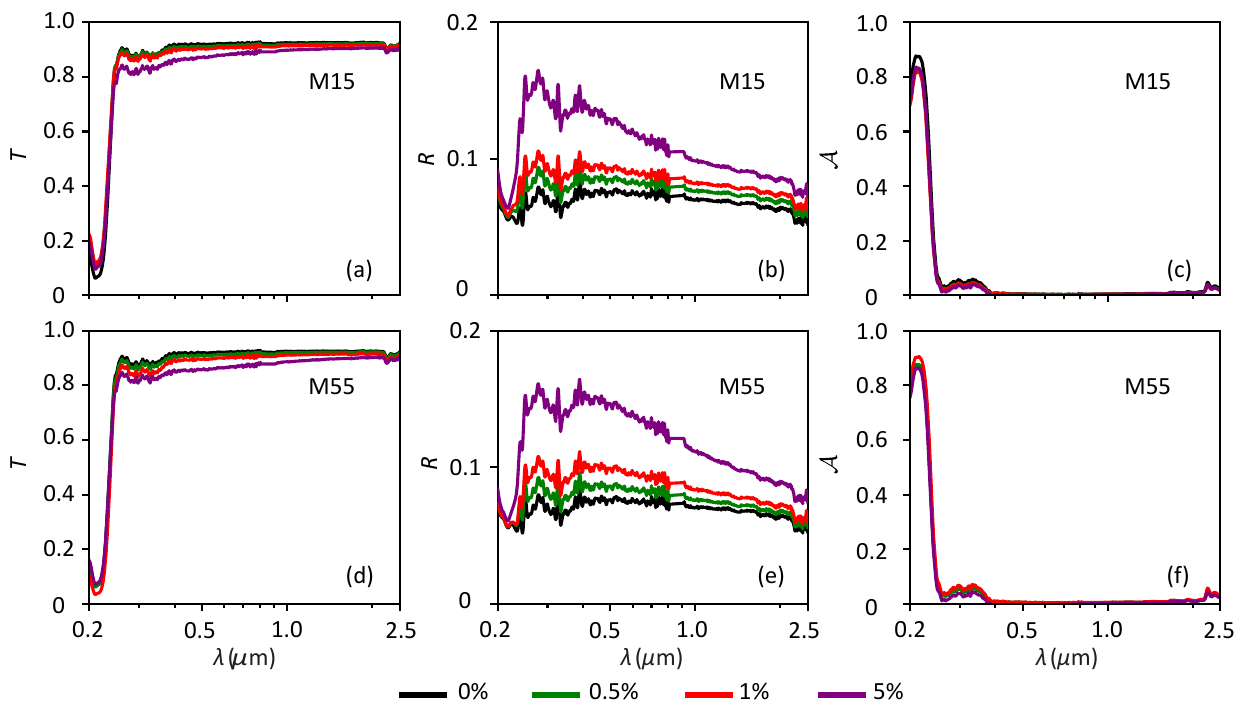}
 \caption{\textbf{(Color online) Prediction results of the overall optical properties of thin-film composites (air bubble/PMMA)}: the total transmittance $T$ in (a),(d), the total reflectance $R$ in (b),(e), and the absorption $\mathcal{A}$ in (c),(f) of series M15 and series M55, respectively. It is assumed that the input data, including sample thickness, volume concentrations, and particle sizes, are the same as those of the TiO$_2$ sample in both series.} \label{fig:RTsim_air}
\end{figure}

Using the validated model to predict the reflectance and  the absorption of thin-film composite in these ranges, we replace TiO$_2$ nanoparticles in the PMMA medium with air bubbles due to their non-absorbing properties. The input data, such as the particle sizes, volume concentrations, and sample thickness, are assumed to be the same as the TiO$_2$ sample series M15 and series M55. As shown in Figs.~\ref{fig:RTsim_air}a and ~\ref{fig:RTsim_air}d, the total transmittance $T$ gradually decreases in comparison with PMMA when increasing the volume concentrations of air bubbles.
Especially, in the wide range of $\lambda \sim$ $0.24-2.5$ $\mu$m, the total reflectance $R$ is improved as shown in Figs.~\ref{fig:RTsim_air}b and~\ref{fig:RTsim_air}e, and the absorption of both series is nearly unchanged with the increasing volume of air in the composite, Figs.~\ref{fig:RTsim_air}c and~\ref{fig:RTsim_air}f. The slight difference between the absorptions of different thin-films is due to the difference in the film thickness.
The wider range of reflectance and the weaker absorption of thin-film composite with air bubbles than those with TiO$_2$ shows the potential of using other components to replace TiO$_2$ toward the application of radiative cooling.
%\phamhong{On the other hand, to minimize the absorption of UV light in the wavelength range of 0.2 to 0.24 $\mu$m, we can replace the absorbing medium with a non-absorbing one. This non-absorbing medium can improve the performance of materials designed for UV applications and also offer enhanced durability and reliability. 
%}

%Our Python-based code \cite{...} provides a simple method to establish this link between film composition and its optical properties.

\section{Conclusion} \label{sec:conclusion}
In this paper, we establish a model based on both Kubelka-Munk theory and Mie theory that links the absorption and scattering properties of individual particles with the reflectance and transmittance spectra of its thin-film composite. With the theories, we study the absorption $K$ and scattering $S$ coefficients of TiO$_2$ nanoparticles immersed in PMMA medium using both experimental and theoretical methods. Thin-film composites are fabricated by spin-coating and then peeled off to form standalone samples for the spectroscopy measurement. The overall optical properties of thin-film composites are measured using the UV-Vis-NIR spectrophotometer in the wavelength range from 0.2 to 2.5 $\mu$m. Using the Kubelka-Munk theory combined with Saunderson correction, we successfully extract the $K$ and $S$ coefficients from the total transmittance $T$ and total reflectance $R$. In the theoretical calculation, we validate the theoretical model in Ref.~\cite{Quinten1996} to determine the absorbing and scattering cross-sections of the nanoparticles immersed in an absorbing medium within Mie theory in the two scenarios: $\mathcal{R}_{\rm{cs}} = d/2$ and $\mathcal{R}_{\rm{cs}} = a$. Our analysis reveals that the agreement between the experimental and theoretical results is better for the first scenario ($\mathcal{R}_{\rm{cs}} = d/2$) than for the second scenario ($\mathcal{R}_{\rm{cs}} = a$). This indicates that the surrounding medium strongly attenuates the absorption and scattering of each particle. By using the theoretical values of $K$ and $S$, we recalculate the overall optical properties of the thin-film composites. The theoretical results show good agreement in the scale and tendency of the experimental data.

The validated theory of absorption and scattering of nanoparticles immersed in an absorbing medium is used to predict the optical properties of other thin-film composites, i.e., air bubbles dispersed in the PMMA medium. From this, we can outlook using the established model of nanocomposites for fast scanning the optical properties before real experiments, making it a useful tool for designing radiative cooling materials. 

%%%%%%%%%%%%%%%%%%%%%%% References %%%%%%%%%%%%%%%%%%%%%%%%%

\section*{Acknowledgments.}
This research is funded by Vietnam National Foundation for Science and Technology Development (NAFOSTED) under grant number 103.02-2021.95. Samples are fabricated in the cleanroom of the Nano and Energy Center. We would like to thank Dr. Pham Do Chung for help with optical measurements. Optical data are measured at VNU University of Science. Sample thicknesses are measured at Jeonbuk National University. {The calculations in this work are done at the PHENIKAA University's HPC Systems.}

\section*{Disclosures.} The author declare no conflicts of interest.
During the preparation of this work the author(s) used ChatGPT and Grammarly in order to improve language and readability. After using this tool/service, the author(s) reviewed and edited the content as needed and take(s) full responsibility for the content of the publication.
%{\section*{Data availability.} Data underlying the results presented in this paper are available}
%in Ref.~\cite{link}.}

%\bibstyle{unsrt}
%\bibliography{composite}

\pagebreak
\clearpage
\appendix
\section{Supplementary Information}

%\begin{center}
%\textbf{\large{Supplementary Material on "Absorption and scattering properties of nanoparticles in an absorbing medium: revisiting with experimental validation"}}
%\end{center}

%\setcounter{section}{0}
\setcounter{equation}{0}
\setcounter{figure}{0}
\setcounter{table}{0}
\setcounter{subsection}{0}
\makeatletter
\renewcommand{\theequation}{A\arabic{equation}}
\renewcommand{\thefigure}{A\arabic{figure}}
\renewcommand{\theHfigure}{A\arabic{figure}}% Hyperref figure hyperlink hook
\renewcommand{\thesubsection}{A\arabic{subsection}}
\subsection{Technical Data of TiO$_2$ Particles}
\label{app1}
\begin{figure}[htb]
 \centering
    \includegraphics[width=0.8\textwidth]{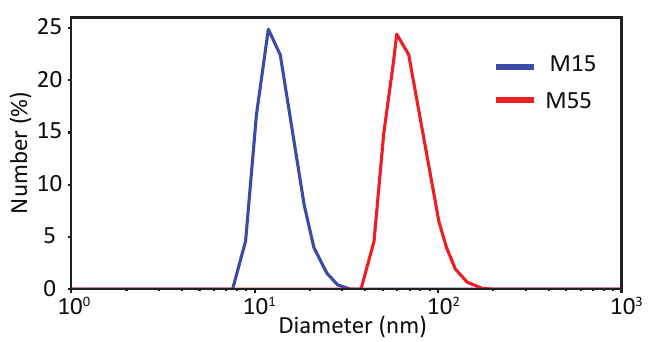}
    \caption{\textbf{Information of TiO$_2$ nanoparticles:} the particle size distribution of TiO$_2$ nanoparticles of the two particle series measured by dynamic light scattering.}
      \label{fig:distribution}
\end{figure}

The size of TiO$_2$ nanoparticles in this research is measured with a dynamic light scattering apparatus using Zeta sizer Nano S equipment from Malvern. TiO$_2$ nanoparticles are dispersed in a water medium by ultrasonic vibration within 4 hours at 25$^\circ$C. After that, samples are passed through a filter membrane with a pore size of 0.45 $\mu$m to remove dust particles or lumps. We perform three continuous measurements to collect the average diameter spectra of TiO$_2$ nanoparticles of two different series, as presented in the blue line and the red line in Fig.~\ref{fig:distribution}. We use a log-normal function to describe the size distribution of TiO$_2$ nanoparticles expressed by
\begin{align}
     \rho(a) = \frac{1}{a\sigma\sqrt{2\pi}} {\rm{exp}}(-\frac{({\rm{ln}} a-\mu)^2}{2\sigma^2}).
\end{align}
Here, $a$ is the particle radius (nm), $\mu$, and $\sigma$  are determined from the values of $\rm{mean}$ and $\rm{mode}$, given by
\begin{align}
	\mu &= 1/3[\ln(\rm{mode}) + 2\ln(\rm{mean})], \\
	\sigma^2 &=2/3[\ln(\rm{mean}) - \ln(\rm{mode})].
\end{align}
Both the $\rm{mean}$ and $\rm{mode}$ values are the experimental results of the particle size measurements, as presented in table~\ref{tab:distribution}. The samples designated M15 and M55 have the average radii of 15 nm and 55 nm, respectively, corresponding to the mean values of the measured diameter of about 30 nm and 110 nm. The parameters in the log-normal function are used as input data to determine the absorption and scattering coefficients of the total particles.

%%%%%%%%%%%%%%%%%%%%%
\subsection{Reflection at Front and Back Interfaces of Thin-Film Composites}
\label{app2}
\begin{figure}[htb]
 \centering
    \includegraphics[width=0.8\textwidth]{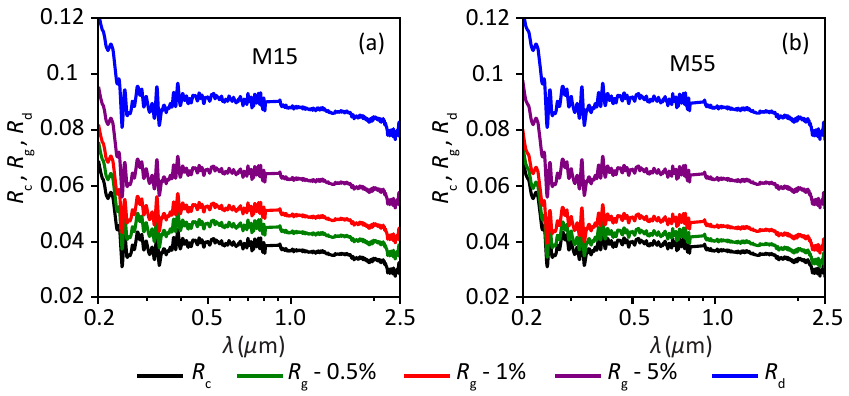}
     \caption{\textbf{Collimated and diffused reflectance:} series M15 in (a), and series M55 in (b). $R_{\rm{c}}$ is the specular reflectance of collimated light at the front interface in the black line, while $R_{\rm{d}}$ in the blue line indicates the reflectance of diffused light calculated by equation~\eqref{eq:Rd}. The backside internal reflectance $R_{\rm{g}}$ is determined by equation~\eqref{eq:Rg} with three mass concentrations: 0.5\% in green line, 1\% in red line, and 5\% in purple line.   }
      \label{fig:RG}
\end{figure}

When light passes through thin-film composites, it undergoes reflection at both the front and back interface of the film. Determining these reflections is necessary to extract the $K-S$ coefficients within the two-flux Kubelka-Munk theory, using equations~\eqref{eq:Tkm}-\eqref{eq:R}. The internal reflectance of the film on the front side is denoted by $R_{\rm{j}}$, while the internal reflectance on the back side is represented by $R_{\rm{g}}$. They are dependent on the fraction of diffused light $q_{\rm{j}}$ and $q_{\rm{g}}$ \cite{Wang18, WANG2020}, expressed by
\begin{align}
      R_{\rm{j}} & = (1-q_{\rm{j}})R_{\rm{c}} + q_{\rm{j}} R_{\rm{d}}, \label{eq:Rj} \\
      R_{\rm{g}} & = (1-q_{\rm{g}})R_{\rm{c}} + q_{\rm{g}} R_{\rm{d}}, \label{eq:Rg}
\end{align}
in which $R_{\rm{c}}$ indicates the fraction of collimated light reflected at the front interface, while $R_{\rm{d}}$ is the diffused reflectance. 
In our thin-film composite, the size of the investigated nanoparticle is smaller than the wavelength of light, so a light scattering profile is considered in both forward and backward directions at angles smaller than the critical angle \cite{Wang18}. The diffused reflectance $R_{\rm{d}}$ is determined using the critical approximation,
\begin{align}
      R_{\rm{d}} & = \frac{\int_{0}^{\theta_{\rm{c}}}{R(\theta)\sin(\theta)\cos(\theta){\rm{d}}\theta}}{\int_{0}^{\theta_{\rm{c}}}{\sin(\theta)\cos(\theta){\rm{d}}\theta}}, \label{eq:Rd} 
\end{align}
in which $\theta_{\rm{c}} = \arcsin(1/n_{\rm{h}})$ is the wavelength-dependent critical angle, $n_{\rm{h}}$ is refractive index of PMMA. $R(\theta)$ is the reflectance of the PMMA medium at the front internal interfaces depending on the incident angle \cite{Howell20@chapter14}, given by
\begin{align}
    R_\theta&=\frac{ R_{\parallel}(\theta) + R_{\perp}(\theta)}{2}, \label{eq:r_theta} \\
    R_{\parallel}(\theta) &= \frac{\cos(\theta)/\cos(\chi) - (n_{\rm{h}} - i\kappa_{\rm{h}})/(n_{\rm{air}} - i\kappa_{\rm{air}})}{\cos(\theta)/\cos(\chi) + (n_{\rm{h}} - i\kappa_{\rm{h}})/(n_{\rm{air}} - i\kappa_{\rm{air}})},  \label{eq:r_perp} \\
    R_{\perp}(\theta) &= \frac{\cos(\chi)/\cos(\theta) - (n_{\rm{h}} -i\kappa_{\rm{h}}) / (n_{\rm{air}} - i\kappa_{\rm{air}})}{\cos(\chi)/\cos(\theta) + (n_{\rm{h}} -i\kappa_{\rm{h}})/(n_{\rm{air}} - i\kappa_{\rm{air}})} \label{eq:r_para}
\end{align}
in which $\theta$ is the incident angle, while $\chi$ is the refraction angle. $(n_{\rm{h}}, \kappa_{\rm{h}})$ and $(n_{\rm{air}}, \kappa_{\rm{air}})$ are the refractive index of PMMA medium and air medium, respectively. 
Fig.~\ref{fig:RG} shows the collimated reflectance $R_{\rm{c}}$ in the black line determined in our previous report \cite{Hong2023} and diffused reflectance $R_{\rm{d}}$ in the blue line calculated by equations~\eqref{eq:Rd}-\eqref{eq:r_para}. From $R_{\rm{c}}$, $R_{\rm{d}}$, and $q_{\rm{g}}$ in table~\ref{tab:q}, the backside internal reflectance $R_{\rm{g}}$ are determined by equation~\eqref{eq:Rg} with three concentrations: 0.5\% in green, 1\% in red, and 5\% in purple, shown in Fig.~\ref{fig:RG}a of series M15 and Fig.~\ref{fig:RG}b of series M55. $R_{\rm{g}}$ increases following the mass concentration, that ranges from $R_{\rm{c}}$ to $R_{\rm{d}}$. 
%%%%%%%%%%%%%%%%%%%%%%%%%5
\subsection{Investigation of the Fractions of Diffused Light  $q_{\rm{j,g}}$}
\label{app3} 
\begin{figure}[htb]
 \centering
    \includegraphics[width=0.8\textwidth]{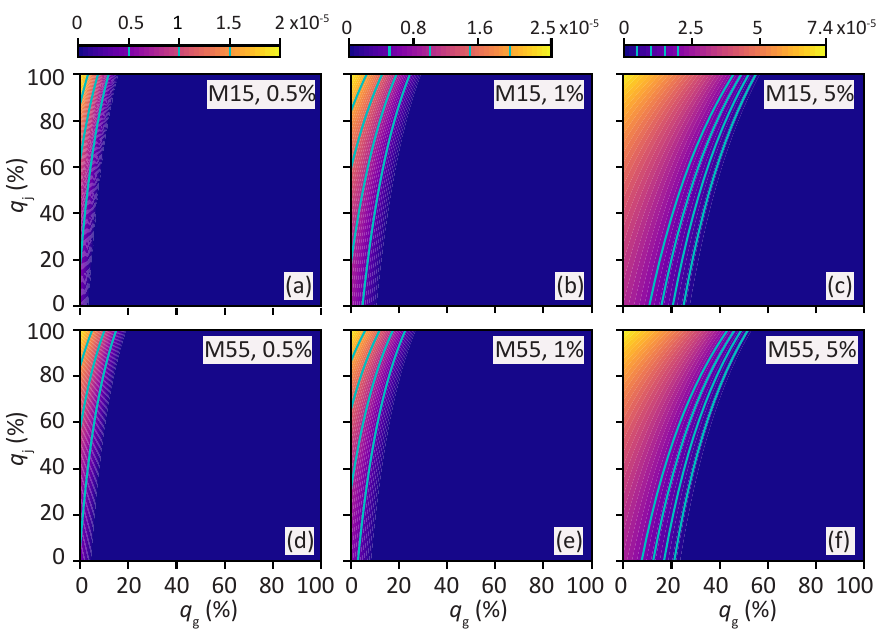}
     \caption{\textbf{Heatmap of the $(K_{\rm{exp}} + S_{\rm{exp}})$ magnitudes in the range of $q_{\rm{j,g}}$ values:} series M15 in (a-c), series M55 in (d-e) with three different concentrations: 0.5\%, 1\%, and 5\%. The values of $q_{\rm{j,g}}$ are valid  when the magnitudes of $(K_{\text{exp}} + S_{\text{exp}})$ are non-negative. } \label{fig:heatmap}
\end{figure}

The fractions of diffuse light at the frontside and backside interfaces, $q_{\rm{j}}$ and $q_{\rm{g}}$, are necessary to determine $R_{\rm{j}}$ and $R_{\rm{g}}$, Eqs.~\ref{eq:Rj}-\ref{eq:Rg}, which are consequently used in extracting $K-S$ coefficients from experimental data of transmittance $T$ and reflectance $R$ within equations~\eqref{eq:Tkm}-\eqref{eq:R}.
However, the two parameters can not be measured directly in this research. 
Therefore, in this appendix, we examine the possible values of $q_{\rm{j,g}}$ by extracting the total $(K_{\rm{exp}} + S_{\rm{exp}})$ from the measured $T$ and $R$ at $\lambda = 0.5$ $\mu m$.
Figure~\ref{fig:heatmap} shows $(K_{\rm{exp}} + S_{\rm{exp}})$ at the varying values of $q_g$ and $q_j$ within $0-100$\% for the series M15 in the top row and the series M55 in the bottom row with the three mass concentrations. All the negative values of $(K_{\rm{exp}} + S_{\rm{exp}})$ are represented by the dark areas (zero value in the color bar). Therefore, with the restriction of non-negative value of $(K+S)$, the possible values of $q_{\rm{j,g}}$ are shown by the light areas in the figures. 
We note that the possible value of $q^{\rm{max}}_{\rm{j}}$ is greater than $2\times q^{\rm{max}}_{\rm{g}}$, it is reasonable as the light 
going from the front to back interface and back has more chance to be scattered by a particle. Therefore, we presume that $q_{\rm{j}}$ $\approx$ $2q_{\rm{g}}$. Since $q_{\rm{j}}\leq 100\%$, then we set $q_{\rm{g}}\leq 50\%$.
Increasing the mass concentration of TiO$_2$ nanoparticles in the composites makes the expansion of the light areas representing the wider range of possible values of $q_{\rm{g}}$; therefore, we assume the fraction of light diffused by particle $q_{\rm{g}}$ equals to the total cross-sections of all the particles in a unit of the sample area,
\begin{align}
    q_{\rm{g}} = N_{\rm{p}}d\int{\pi a^2 \rho(a){\rm{d}}a},
\end{align}
in which $N_{\rm{p}}$ is the particle density, expressed by equation~\eqref{eq:Np}, and $d$ is sample thickness.
\subsection{Effective Refractive Index of the Particulate Medium}
\label{app4}
\begin{figure}[htb]
 \centering
    \includegraphics[width=0.8\textwidth]{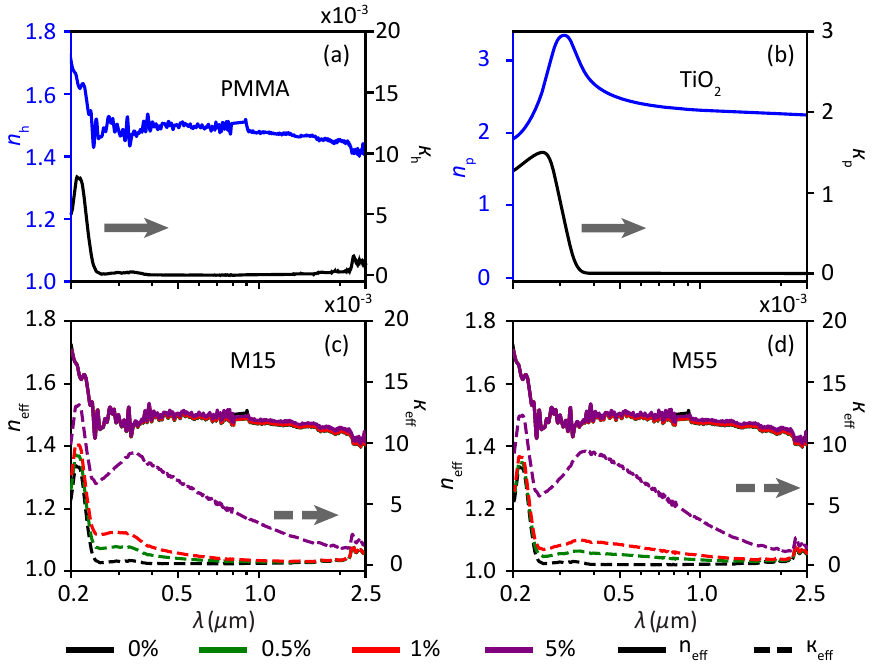}
     \caption{\textbf{Complex refractive index:} (a) refractive index and extinction coefficient of PMMA in our previous research \cite{Hong2023}, and (b) TiO$_2$ adapted from Ref.~\cite{Siefke2016}. The particulate medium with refractive index $n_{\rm{eff}}$ in solid line and extinction coefficient $\kappa_{\rm{eff}}$ are presented in (c) of series M15 and in (d) of series M55 with different concentrations.}
      \label{fig:nk_eff}
\end{figure}
The refractive index of PMMA $(n_{\rm{h}},\kappa_{\rm{h}})$ and TiO$_2$ particle $(n_{\rm{p}},\kappa_{\rm{p}})$ are shown in Fig.~\ref{fig:nk_eff}a and Fig.~\ref{fig:nk_eff}b, respectively. The optical properties of PMMA are studied in our previous research \cite{Hong2023}, while those of TiO$_2$ are adapted from Ref.~\cite{Siefke2016}.
%The optical properties of PMMA are shown in Fig.~\ref{fig:nk_eff}a, which is studied in our previous research~\cite{Hong2023}. Fig.~\ref{fig:nk_eff}b is the refractive index of TiO$_2$ adapted from Ref.~\cite{Siefke2016}.
When TiO$_2$ particles disperse in PMMA medium, the thin-film composites become a particulate medium with effective refractive index ($n_{\rm{eff}}$, $\kappa_{\rm{eff}}$). It represents the mean effect of other particles inside the composite on a given particle, which is expressed through the effective permittivity $\varepsilon_{\rm{eff}}$ and permeability $\mu_{\rm{eff}}$ \cite{Wheeler09, William1997},
\begin{align}
    n_{\rm{eff}} + i\kappa_{\rm{eff}} & = \sqrt{\varepsilon_{\rm{eff}} \mu_{\rm{eff}}}. 
\end{align}
The effective permittivity $\varepsilon_{\rm{eff}}$ and permeability $\mu_{\rm{eff}}$ are associated with the forward and backward scattering amplitudes of particles hosted in the matrix, $S(a,0)$ and $S(a,\pi)$ \cite{Bohren83}. They are given by
\begin{align}
    \varepsilon_{\rm{eff}} & = \varepsilon_{\rm{0}} \frac{1 + (2/3)\gamma}{1-\gamma/3}, \label{eq:eeff} \\ 
    \mu_{\rm{eff}} & = \frac{1 + (2/3)\delta}{1-\delta/3}, \label{eq:meff}
\end{align}
here $\varepsilon_{\rm{0}}$ is the permittivity of the PMMA matrix, $\varepsilon_{\rm{0}} = (n_{\rm{h}} + \kappa_{\rm{h}})^2$, and
\begin{align} 
     \gamma & = \int\frac{3i}{2 \bar{x}^3}\left[N_{\rm{p}}\frac{4\pi}{3}a^3 \rho(a)\right][S(a,0)+S(a,\pi)] {\rm{d}}a, \\ 
     \delta & = \int\frac{3i}{2 \bar{x}^3}\left[N_{\rm{p}}\frac{4\pi}{3}a^3 \rho(a)\right][S(a,0)-S(a,\pi)] {\rm{d}}a, 
\end{align}
where $N_{\rm{p}}$ is the particle density, $a$ is the particle radius, $\rho(a)$ is the particle size distribution, $\bar{x} = 2\pi(n_{\rm{h}} + i \kappa_{\rm{h}})a/\lambda$ is the size parameter.

Fig.~\ref{fig:nk_eff}c and Fig.~\ref{fig:nk_eff}d show the refractive indices of particular media of series M15 and series M55 with four concentrations: 0\% in black, 0.5\% in green, 1\% in red, and 5\% in purple. All the lines representing the effective refractive index $n_{\rm{eff}}$ of the particulate medium almost overlap with that of PMMA. It means that the refractive index is weakly affected by particles in all samples. In contrast, the effective extinction coefficient $\kappa_{\rm{eff}}$ shown in the dashed line increases as the TiO$_2$ particle concentrations increase. 

\clearpage

\bibstyle{unsrt}
\bibliography{composite}

\end{document}